\documentstyle[12pt]{article}
\def\II{{I}}
\def\RR{{\bf R}}
\def\CC{{\bf C}}

\def\tr{{\rm tr\,}}
\def\Tr{{\rm Tr\,}}

\def\det{{\rm det\,}}

\def\End{{\rm End\,}}

\def\log{{\rm log\,}}

\def\rank{{\rm rank\,}}

\def\vol{{\rm vol\,}}

\def\be{\begin{equation}}
\def\ee{\end{equation}}
\def\bea{\begin{eqnarray}}
\def\eea{\end{eqnarray}}
\newtheorem{theorem}{Theorem}

\def\uiowano#1{{
\null\vskip-3truecm
\hspace*{6truecm}{\hrulefill}\par\vskip-4truemm\par
\hspace*{6truecm}{\hrulefill}\par\vskip5mm\par
\hspace*{6truecm}{{\large\sc The University of Iowa
{\rm (#1)}}}\vskip4mm\par
\hspace*{6truecm}{\hrulefill}\par\vskip-4truemm\par
\hspace*{6truecm}{\hrulefill}
\par\bigskip\par
}}

\begin{document}

\begin{titlepage}
\uiowano{Dec. 1998}
\begin{minipage}{6truecm}
\ \end{minipage}
\begin{minipage}{9truecm}
{\large\sf DSF 98/44}\\ \\
{\sf to appear in:
Proc. of the Conf. ``Trends in Mathematical Physics'',
Knoxville, October 14--17, 1998, \\
(Cambridge: International Press (1999))
}
\end{minipage}

\par
\vfill

\centerline{\LARGE\bf Heat Kernel Asymptotics}
\medskip
\centerline{\LARGE\bf of Gilkey--Smith
Boundary-Value Problem}
\bigskip\bigskip
\centerline{\Large\bf Ivan G. Avramidi
\footnote{On leave of absence from Research Institute for Physics, 
Rostov State University,  Stachki 194, 344104 Rostov-on-Don, Russia.}}
\bigskip
\centerline{\it Department of Mathematics, The University of Iowa}
\centerline{\it 14 MacLean Hall, Iowa City, IA 52242-1419, USA}
\centerline{\it E-mail: iavramid@math.uiowa.edu} 
\smallskip
\centerline{and}
\smallskip
\centerline{\Large\bf Giampiero Esposito}
\bigskip
\centerline{\it Istituto Nazionale di Fisica Nucleare, Sezione di Napoli}
\centerline{\it Mostra d'Oltremare Padiglione 20, 80125 Napoli, Italy}
\centerline{\it E-mail: giampiero.esposito@na.infn.it}

\vfill

{\narrower\par
The formulation of gauge theories on compact Riemannian manifolds with boundary
leads to partial differential operators with Gilkey--Smith boundary conditions,
whose peculiar property is the occurrence of both normal and tangential
derivatives on the boundary.  Unlike the standard Dirichlet or Neumann boundary
conditions, this boundary-value problem is not automatically elliptic but
becomes elliptic under certain conditions on the boundary operator.  We study
the Gilkey--Smith boundary-value problem for Laplace-type operators and find a
simple criterion of ellipticity.  The first non-trivial coefficient of the
asymptotic expansion of the trace of the heat kernel is computed and the local
leading asymptotics of the heat-kernel diagonal is also obtained.  It is shown
that, in the non-elliptic case, the heat-kernel diagonal is non-integrable near
the boundary, which reflects the fact that the heat kernel is not of trace
class.  We apply this analysis to general linear bosonic gauge theories and find
an explicit condition of ellipticity.

\par}
\vfill
\end{titlepage}


\section{Introduction}

Elliptic differential operators on manifolds 
have proved to play a crucial role in
mathematical physics.  In particular, the main objects of interest in quantum
field theory and statistical physics, such as the effective action and the
partition function, are described by the functional determinants, or, which is
equivalent, by the zeta-function and the heat kernel of self-adjoint elliptic
differential operators.  Of particular importance are, of course, the
operators of Laplace type or Dirac type.\cite{avresp-cmp98}
In the case of manifolds with boundary, one has to impose some
boundary conditions in order to make a (formally self-adjoint) differential
operator self-adjoint and elliptic.  
Indeed, the boundary conditions are additional ingredients
in the theory which have not been fixed {\it a priori}, and the choice of
boundary conditions is, by no means, unique. There are many admissible
boundary conditions that guarantee the
self-adjointness and ellipticity of the problem. The simplest boundary
conditions are the classical Dirichlet and the Neumann ones. In the Dirichlet
case one sets to zero at the boundary the value of the field, whereas in the
Neumann case the normal derivative of the field is set to zero at the
boundary. There exist also slight modifications of the Neumann boundary
conditions (called Robin boundary conditions in physical literature) when the
normal derivative of the field at the boundary is not set to zero but is
proportional to the value of the field at the boundary.  An even more general
scheme, called mixed boundary conditions, applies to the operators acting on
sections of some vector bundles. It is then possible to mix the
Dirichlet and Robin boundary conditions by using some projectors, i.e.  a
part of the field components satisfy Dirichlet boundary conditions and
the remaining part satisfies Robin boundary conditions.

However, this is not the most general scheme. A much more general setup for the
boundary-value problem was developed by Gilkey and 
Smith.\cite{gilkeysmith83b} They put 
forward some boundary conditions that are
still local but include both normal and {\it tangential derivatives}
of the fields at the boundary.  In this paper we are going to study the
Gilkey--Smith boundary-value problem for operators of Laplace type. 
Unlike the Dirichlet or Neumann boundary-value problems, such
a boundary-value problem is not automatically elliptic. Therefore, following
Refs. 1,3, we find, first, a criterion of
(strong) ellipticity, which provides an explicit simple condition on the
boundary operator.  Then we construct the parametrix to the heat equation in
the leading approximation and compute the first non-trivial (next to 
leading) term $A_{1/2}$ in the asymptotic expansion of the trace of the heat
kernel.  We also discuss what happens when the boundary-value problem is not
strongly elliptic.  Last, we study the problem of ellipticity in linearized
gauge theories on manifolds with boundary.  The attempt to preserve gauge
invariance on manifolds with boundary fixes the boundary conditions and leads
exactly to a Gilkey--Smith boundary-value problem. 
As is shown in Ref. 1, while Yang--Mills as well as Rarita--Schwinger
theories are automatically elliptic, quantum gravity based on the Einstein
action turns out to be not elliptic, if the Gilkey--Smith 
boundary-value problem is studied.

\section{Gilkey--Smith Boundary-Value Problem}

Let $(M,g)$ be a smooth compact Riemannian manifold of dimension $m$ with
smooth boundary $\partial M$.  Let $g$ be the positive-definite Riemannian
metric on $M$ and $\hat g$ be the induced metric on $\partial M$.  By using
the inward geodesic flow, we identify a narrow neighbourhood of the boundary
$\partial M$ with a part of $\partial M\times \RR_+$ and define a split of the
cotangent bundle $T^*(M) =T^*(\partial M) 
\oplus T^*(\RR)$.  Let $\hat x=({\hat x}^{i})$, with
$i=1,2,\dots,m-1$, be the local coordinates on $\partial M$ and $r$
be the normal geodesic distance to the boundary, so that $N=\partial_r=
\partial/\partial r$ is the inward-pointing unit normal vector field to the
boundary.  Near $\partial M$ we choose the local coordinates $x=(x^\mu)=(\hat
x,r)$, with $\mu=1,2,\dots,m$, and the split 
$\xi=(\xi_\mu)=(\zeta,\omega) \in T^{*}(M)$,
where $\zeta=(\zeta_j) \in T^*(\partial M)$ and $\omega\in \RR$.  With
our notation, Greek indices range from 1 through $m$ and lower case Latin
indices range from 1 through $m-1$.

Let $V$ be a (smooth) vector bundle over the manifold $M$ and $C^\infty(V,M)$
be the space of smooth sections of the bundle $V$.  Let $V^*$ be the dual
vector bundle and $E:\ V\to V^*$ be a Hermitian non-degenerate metric, $
E^{\dag}=E $, that determines the Hermitian fibre scalar product in $V$.
Using the invariant Riemannian volume element 
${\rm dvol}(x)$ on $M$ one defines a
natural $L^2$ inner product $( , )$ in $C^\infty(V,M)$, and the Hilbert space
$L^2(V,M)$ as the completion of $C^\infty(V,M)$ in this norm.

Let $\nabla^{T^*(M)}$ be the Levi-Civita connection on $M$ and $\nabla^V$ be
the connection on the vector bundle $V$ compatible with the metric $E$.  Then
we define, as usual, $ \nabla^{T^*(M) \otimes V}=\nabla^{T^*(M)}\otimes
1+1\otimes\nabla^{V}$. Moreover, we will often denote just by $\nabla$ the
total covariant derivative without mentioning the bundle it is acting on.  
The notation $\hat\nabla$ will be used to denote the covariant tangential
derivative on the boundary.

Let, further, $\tr_g=g\otimes 1$ be the contraction of sections of the bundle
$T^*(M) \otimes T^*(M) \otimes V$ with the 
metric on the cotangent bundle $T^*(M)$,
and $Q\in C^\infty(\End(V),M)$ be a smooth self-adjoint endomorphism of the
bundle $V$, i.e.  $\bar Q\equiv E^{-1} Q^{\dag}E=Q $.  Then a Laplace-type
operator $F:\ C^\infty(V,M)\to C^\infty(V,M)$ is a second-order
differential operator defined by
\be 
F \equiv
-\tr_g\nabla^{T^*(M) \otimes V}\nabla^V+Q.  
\label{1} 
\ee
Let us define the {\it boundary data}
by 
\be
\psi(\varphi)=\left(\matrix{\psi_0(\varphi)\cr
\psi_1(\varphi)\cr}\right),
\label{2}
\ee
where $ \psi_0(\varphi)\equiv\varphi|_{\partial M}$ and 
$\psi_1(\varphi)\equiv\nabla_N\varphi|_{\partial M}$ are the restrictions 
to the boundary of the
sections $\varphi\in C^\infty(V,M)$ and their normal derivatives. 
Let the vector bundle $W$ over $\partial M$ be the bundle of the
boundary data.  $W$ consists of two copies of the restriction of $V$ to
$\partial M$ and inherits a natural grading\cite{gilkey95} $W=W_0\oplus W_1$,
where $W_j$ represents normal derivatives of order $j$, and, therefore, $ \dim
W=2\dim V.  $ The bundles $W_0$ and $W_1$ have the same structure, and hence
in the following they will be often identified.  Let $W'=W'_0\oplus W'_1$ be
an auxiliary graded vector bundle over $\partial M$ such that $ \dim W'=\dim
V.  $ Let $B:  C^\infty(W,\partial M)\to C^\infty(W',\partial M)$ be a
tangential differential operator on $\partial M$.  The boundary conditions
then read
\be
B\psi(\varphi)=0.
\label{3}
\ee

Now let $\Pi$ be a self-adjoint projector acting on $W_0$.  Since the bundle
$W_{1}$ is identified with $W_0$ the projector $\Pi$ acts also on $W_1$.  Let
$\Gamma\in C^\infty(T(\partial M) \otimes {\rm End}\,(W_0),\partial M)$ be an
anti-self-adjoint endomorphism-valued vector field on $\partial M$ orthogonal
to $\Pi$, i.e.  $\bar\Gamma^i=-\Gamma^i$, $\Pi\Gamma^i=\Gamma^i\Pi=0$, and
$S\in C^\infty(\End(W_0),\partial M)$ be a self-adjoint endomorphism 
orthogonal to $\Pi$,
i.e.  $\bar S=S$, $\Pi S=S\Pi=0$.  Using these objects we define a first-order
self-adjoint tangential differential operator $\Lambda:\ C^\infty(W_0,\partial
M)\to C^\infty(W_0,\partial M)$ by
\be
\Lambda=(\II-\Pi)\left\{
{1\over 2}(\Gamma^i\hat\nabla_i+\hat\nabla_i\Gamma^i)
+S\right\}(\II-\Pi),
\label{4}
\ee
that is obviously orthogonal to $\Pi$:  $\Pi\Lambda=\Lambda\Pi=0$. 
Hereafter $\II$ is the identity endomorphism of the vector bundle $V$.
The Gilkey--Smith boundary operator is expressed in terms of 
these geometric objects by
\be
B=\left(\matrix{\Pi&0\cr
\Lambda& \II -\Pi\cr}\right).
\label{5}
\ee
It is not difficult to see that the Gilkey--Smith boundary-value problem
incorporates all standard types of boundary conditions. Indeed, by choosing
$\Pi=\II$ and $\Lambda=\Gamma=S=0$ one gets the standard Dirichlet boundary
conditions, by choosing $\Pi=0$, $\Gamma=0$, $\Lambda=S=\II$ one gets the
standard Neumann boundary conditions.  More generally, the choice $\Gamma=0$
and $\Lambda=S$ corresponds to the mixed boundary conditions
mentioned in Sec. 1.

\section{Strong Ellipticity}

Integration by parts shows\cite{avresp-cmp98} that the Laplace-type
operator $F$ given in Eq. (\ref{1}) 
endowed with the Gilkey--Smith boundary conditions
(\ref{3}) is symmetric, meaning that $(\varphi_1,F\varphi_2) =
(F\varphi_1,\varphi_2)$ for any two smooth sections $\varphi_1, \varphi_2 \in
C^\infty(V,M)$ satisfying the boundary conditions $B\psi(\varphi_1) =
B\psi(\varphi_2) = 0$.  However, it is not automatically elliptic.  Now we are
going to determine under which conditions the Gilkey--Smith boundary-value
problem for a Laplace-type operator is strongly elliptic.\cite{gilkey95}

First of all, the leading symbol of the operator $F$ should be elliptic in the
interior of $M$.  Let, hereafter, $\lambda$ be a complex number which does not
lie on the positive real axis, $\lambda\in {\bf C}-{\bf R}_+$ (${\bf R}_+$
being the set of positive numbers). Then by using the leading
symbol of the operator $F$, i.e. $\sigma_L(F;x,\xi)=|\xi|^2\cdot
\II$, with $|\xi|^2 \equiv g^{\mu\nu}(x)\xi_\mu\xi_\nu $, we find easily
\be
\det (\sigma_L(F;x,\xi)-\lambda\cdot\II)
=(|\xi|^2-\lambda)^{\dim V}\ne 0
\qquad {\rm for}\ (\xi,\lambda)\ne (0,0).
\label{6}
\ee
Thus, the leading symbol of the operator $F$ is {\it elliptic}.

Second, the so-called {\it strong ellipticity condition} should be 
satisfied.\cite{gilkeysmith83b,gilkey95}
As we already noted above, there is a natural grading in the vector bundles
$W$ and $W'$ which reflects simply the number of normal derivatives
of a section of the bundle.\cite{gilkey95}
The boundary operator $B$ in Eq. (\ref{5}) 
is said to have the {\it graded order} $0$.
Its {\it graded leading symbol} is defined 
by\cite{gilkeysmith83b,gilkey95}
\be
\sigma_{g}(B_{F}) \equiv \left(\matrix{\Pi&0\cr
i\Gamma\cdot\zeta& (\II-\Pi)}\right),
\label{7}
\ee
where $\Gamma\cdot\zeta\equiv\Gamma^{j}\zeta_{j}$.
To define the
strong ellipticity condition we take the leading symbol $\sigma_L(F;\hat x, r,
\zeta, \omega)$ of the operator $F$, substitute $r=0$ and $\omega\to
-i\partial_{r}$ and consider the following ordinary differential equation for
a section $\varphi\in C^\infty(V,\partial M\times {\bf R}_+)$:
\be
\left[\sigma_{L}(F;\hat x, 0, 
\zeta, -i\partial_{r})-\lambda\cdot \II\right]\varphi(r)=0,
\label{8}
\ee
with an asymptotic condition
\be
\lim_{r\to\infty}\varphi(r)=0.
\label{9}
\ee

The boundary-value problem $(F,B)$ is said to be 
{\it strongly elliptic}\cite{gilkeysmith83b,gilkey95} 
with respect to the cone $\CC-\RR_{+}$ if for
every pair $(\zeta,\lambda)\ne (0,0)$, and any $\psi'\in C^\infty(W',\partial
M)$ there is a {\it unique} solution $\varphi$ of the equation (\ref{8})
satisfying the asymptotic condition (\ref{9}) and
the boundary condition
\be
\sigma_g(B_{F})({\hat x},\zeta)\psi(\varphi)=\psi' ,
\label{10}
\ee
where $\psi(\varphi)\in C^\infty(W,\partial M)$ are the boundary data 
defined by (\ref{2}).

For a Laplace-type operator this definition leads to the
following theorem.\cite{avresp-cmp98}  
\begin{theorem}

The Gilkey--Smith boundary-value problem $(F,B)$ is strongly elliptic
with respect to $\CC-\RR_{+}$ if and only if the matrix $|\zeta| \II
-i\Gamma\cdot\zeta$ is positive-definite, i.e.
$|\zeta|\II-i\Gamma\cdot\zeta>0$, for any non-vanishing $\zeta$.  
A sufficient condition for strong ellipticity is:  
\be
|\zeta|^2\II+(\Gamma\cdot\zeta)^2>0.
\label{11}
\ee

\end{theorem}

\section{Asymptotic Expansion of the Heat Kernel}

For $t>0$ the heat semi-group $\exp(-tF):\ L^{2}(V,M)\to L^2(V,M)$ of the
strongly elliptic boundary-value problem $(F,B)$ is well defined.  The kernel
of this operator, $U(t|x,y)$, called heat kernel, is a section of the tensor
product of the vector bundles $V$ and $V^*$ over the tensor-product manifold
$M\times M$, defined by the equation
\be
(\partial_t+F)U(t|x,y)=0
\label{12}
\ee
with initial condition
\be
U(0^+|x,y)=\delta(x,y),
\label{13}
\ee
where $\delta(x,y)$ is the covariant Dirac distribution. 
Moreover, a boundary condition is imposed, i.e.
\be
B\psi[U(t|x,y)]=0,
\label{14}
\ee
and the symmetry condition holds
\be
U(t|x,y)=U(t|y,x).
\label{15}
\ee
Hereafter all differential operators as well as the boundary data map act on
the {\it first} argument of the heat kernel, unless otherwise stated.

It is well known\cite{gilkey95} that the heat kernel $U(t|x,y)$ is a smooth
function near diagonal of $M\times M$ and has a well defined diagonal value
$U(t|x,x)$, and that the $L^2$ trace
\be
\Tr_{L^2}\exp(-tF)=\int_M {\rm dvol}(x)\tr_V U(t|x,x),
\label{16}
\ee
has an asymptotic expansion as $t\to 0^+$
\be
\Tr_{L^2}\exp(-tF)\sim (4\pi t)^{-m/2}\sum\limits_{k\ge 0}
t^{k/2}A_{k/2}(F,B).
\label{17}
\ee
Here $A_{k/2}(F,B)$ are the famous {\it global} heat-kernel
coefficients (sometimes called also Minakshisundaram--Plejel or Seeley
coefficients). The zeroth-order coefficient is very well known:
\be
A_0=\int\limits_M {\rm dvol}(x)\tr_V \II=\vol(M)\cdot\dim(V).
\label{18}
\ee
It is independent of the operator $F$ and of the boundary conditions $B$.
The higher order coefficients have the following general form:
\be
A_{k/2}(F,B)=\int\limits_M {\rm dvol}(x)\tr_V a_{k/2}(F|x)
+\int\limits_{\partial M}{\rm dvol}(\hat x)
\tr_V b_{k/2}(F,B|\hat x),
\label{19}
\ee
where $a_{k/2}(F|x)$ and $b_{k/2}(F,B|{\hat x})$ are 
the ({\it local}) {\it interior} and 
{\it boundary} heat-kernel coefficients.  The interior coefficients 
$a_{k/2}(F|x)$ do {\it not}
depend on the boundary conditions. Moreover, it is well known
that they vanish for half-integer order, $a_{k+1/2}=0$.\cite{gilkey95} 
The integer order coefficients $a_k(F|x)$ are 
calculated for Laplace-type operators up
to $a_4$.\cite{avra91b}  The boundary 
coefficients $b_{k/2}(F,B|{\hat x})$ do depend
on {\it both} the operator $F$ and the boundary operator $B$.  They are far
more complicated because in addition to the geometry of the manifold $M$ they
depend essentially on the geometry of the boundary $\partial M$.  For
Laplace-type operators they are known for the usual boundary conditions
(Dirichlet, Neumann, or mixed version of them) 
up to $b_{5/2}$.\cite{branson90,kirsten}
For Gilkey--Smith boundary conditions
only some special cases have been studied in the 
literature.\cite{mcavity91,dowker97,avresp97,dowker98,eli98}
In this paper we evaluate the next-to-leading coefficient
$A_{1/2}(F,B)$, following our recent work.\cite{avresp-cmp98}

\section{Parametrix: General Setup}

In this section we show how one can construct an approximation to
the heat kernel $U(t|x,y)$ for $t\to 0^{+}$ near the diagonal, i.e.  for $x$
close to $y$.
First of all, we decompose the heat kernel into two parts
\be
U(t|x,y)=U_\infty(t|x,y)+U_B(t|x,y).
\label{20}
\ee
Then we construct {\it different} approximations for $U_\infty$ and
$U_B$.  The first part $U_\infty(t|x,y)$ is approximated by the usual
asymptotic expansion of the heat kernel in the case of compact manifolds {\it
without boundary} when $x\to y$ and $t\to 0^{+}$.  This means that effectively
one introduces a small expansion parameter $\varepsilon$ reflecting the fact
that the points $x$ and $y$ are close to each other and the parameter $t$ is
small.  This can be done by fixing a point $x'$, choosing the normal
coordinates at this point (with $g_{\mu\nu}(x')=\delta_{\mu\nu}$) and scaling
\be
x\to x'+\varepsilon(x-x'), \qquad y\to x'+\varepsilon(y-x'), 
\qquad t\to \varepsilon^{2}t,
\label{21}
\ee
and expanding into an asymptotic series in $\varepsilon$.
This construction is, by now, quite standard\cite{gilkey95} 
and we do not repeat it here.
One can also use a completely covariant method,\cite{avra91b}
which leads to the result
\be
U_\infty(t|x,y)\sim (4\pi t)^{-m/2}\exp\left(-{d^2(x,y)\over 4t}\right)
\sum\limits_{k\ge 0} {t^k}a_k(x,y),
\label{22}
\ee
where $d(x,y)$ is the geodesic distance
between $x$ and $y$ and $a_k(x,y)$ are the
off-diagonal heat-kernel coefficients.  These coefficients satisfy certain
differential recursion relations which can be solved in form of a covariant
Taylor series near diagonal.\cite{avra91b}  On the diagonal the asymptotic
expansion of the heat kernel reads
\be
U_\infty(t|x,x)
\sim (4\pi t)^{-m/2}\sum\limits_{k\ge 0}t^{k}a_{k}(F|x),
\label{23}
\ee
where $a_k(F|x)\equiv a_k(x,x)$. As we noted above,
the explicit formulae for the diagonal values of $a_k$ are known up 
to $k=4$.\cite{avra91b} 
This asymptotic expansion can be integrated over
the manifold $M$ to get
\be
\int\limits_M {\rm dvol}(x)\tr_V U_\infty(t|x,x)
\sim (4\pi t)^{-m/2}\sum\limits_{k\ge 0}t^{k}
\int\limits_M {\rm dvol}(x)\tr_V\,a_{k}(F|x).
\label{24}
\ee
Thus, integrating the diagonal of $U_\infty$ gives the 
interior terms in the heat-kernel asymptotics (\ref{19}). 

For a {\it strongly elliptic} boundary-value problem the diagonal of the
boundary part $U_B(t|x,x)$ is {\it exponentially small} as $t\to 0^{+}$ if
$x\not\in\partial M$, i.e. of order 
$\sim\exp(-r^2(x)/t)$, where $r(x)$ is the
normal geodesic distance from $x$ to the 
boundary. Thus, it does not contribute
to the asymptotic expansion of the heat-kernel diagonal outside the 
boundary as $t \to 0^{+}$. This implies that  
the asymptotic expansion of the total heat-kernel
diagonal outside the boundary is determined only by $U_\infty$ 
\be
 U(t|x,x)\sim
(4\pi t)^{-m/2}\sum\limits_{k\ge 0} {t^k}a_k(F|x), 
\qquad x\not\in \partial M.
\label{25}
\ee

The coefficients of the asymptotic expansion 
as $t \to 0^{+}$ of the diagonal of
the boundary part $U_B(t|x,x)$ behave near the boundary like the
one-dimensional Dirac distribution $\delta(r(x))$ and its derivatives.  Thus,
the {\it integral} over the manifold $M$ of the boundary part $U_B(t|x,x)$ has
an asymptotic expansion as $t \to 0^{+}$ with non-vanishing coefficients in form
of integrals over the boundary. The local boundary coefficients $b_{k/2}$
contribute, after integration over the boundary, to the global
heat-kernel coefficients, according to Eq. (\ref{19}).
It is well known that the coefficient $A_{1/2}$ is a purely boundary 
contribution.\cite{gilkey95}  It is almost 
obvious that it can be evaluated by integrating
the fibre trace of the boundary contribution $U_{B}$ of the heat kernel to
leading order.

Of course, $U_\infty$ is obtained {\it without} taking into account the
boundary conditions.  Therefore, it satisfies approximately the equation
(\ref{12}) but does {\it not} satisfy the boundary conditions
(\ref{14}). This implies that the
compensating term $U_B(t|x,y)$ should be defined by the equation
\be
(\partial_t+F)U_B(t|x,y)=0
\label{26}
\ee
with the initial condition
\be
U_B(0^+|x,y)=0,
\label{27}
\ee
and the boundary condition
\be
B\psi\left[U_\infty(t|x,y)
+U_B(t|x,y)\right]=0.
\label{28}
\ee
The compensating term $U_{B}(t|x,y)$ is important only near the boundary where
it behaves like a distribution when 
$t\to 0^{+}$.  Since the points $x$ and $y$
are close to the boundary the coordinates $r(x)$ and $r(y)$ are small {\it
separately}, hence not only the difference $[r(x)-r(y)]$ but also the sum
$[r(x)+r(y)]$ is small.  This means that we must additionally scale 
$r(x)\to\varepsilon r(x)$ and $r(y)\to \varepsilon r(y)$.
By contrast, the point $\hat x'$ is kept fixed on the boundary, so the
coordinates $\hat x'$ do not scale at all:  $\hat x'\to \hat x'$.

Thus, we shall scale the coordinates $x=(\hat x,r(x))$, $y=(\hat y,r(y))$,
and the parameter $t$ according to
\be
\hat x\to \hat x'+\varepsilon(\hat x-\hat x'),\qquad 
\hat y\to \hat x'+\varepsilon(\hat y-\hat x'),
\label{29}
\ee
\be
r(x)\to \varepsilon r(x),\qquad 
r(y)\to \varepsilon r(y), \qquad
t\to \varepsilon^2 t .
\label{30}
\ee
The corresponding differential operators are scaled by
\be
\hat\partial\to{1\over\varepsilon}\hat\partial, \qquad
\partial_r\to{1\over\varepsilon}\partial_r,\qquad
\partial_t\to{1\over\varepsilon^2}\partial_t.
\label{31}
\ee

We call this transformation just {\it scaling} and denote the scaled objects by
an index $\varepsilon$, e.g.  $U_B^{\varepsilon}$.  The scaling parameter
$\varepsilon$ is considered as a small parameter in the theory and we use it to
expand everything in power series in $\varepsilon$.  We do {\it not} take care
about the convergence properties of these expansions and take them as {\it
formal} power series.  In fact, they are asymptotic expansions as
$\varepsilon\to 0$.  At the very end of calculations we can set $\varepsilon=1$.
The non-scaled objects, i.e.  those with $\varepsilon=1$, will not have the
index $\varepsilon$, e.g.  $U_B^\varepsilon|_{\varepsilon=1}=U_B$.  Another way
of doing this is by saying that we expand all quantities in the Taylor series in
the boundary coordinates $\hat x$ and $\hat y$ about the point $\hat x'$ with
the coefficients being homogeneous functions of $r(x)$, $r(y)$ and $t$.

First of all, we expand the scaled operator $F^\varepsilon $ in power 
series in $\varepsilon$ 
\be
F^\varepsilon\sim \sum\limits_{n\ge 0}\varepsilon^{n-2} F_n,
\label{32}
\ee
where $F_n$ are second-order differential operators with homogeneous symbols.
The boundary operator requires a more careful handling.  Since half of the
boundary data (\ref{2}) contain normal derivatives, formally
$\psi_0=\varphi|_{r=0}$ and $\psi_1=\partial_r\varphi|_{r=0}$ would
be of different order in $\varepsilon$.  To make them of the same order we have
to assume an additional factor $\varepsilon$ in all $\psi_1\in
C^{\infty}(W_1,\partial M)$.  Thus, we define the {\it graded scaling} of the
boundary data map by
\be
\psi^\varepsilon(\varphi)=\left(\matrix{\psi^\varepsilon_0(\varphi)\cr
\varepsilon\psi^\varepsilon_1(\varphi)\cr}\right)
=\left(\matrix{\varphi(\hat x, r)|_{r=0}\cr
\partial_r\varphi(\hat x,r)|_{r=0}\cr}\right)=\psi(\varphi),
\label{33}
\ee
so that the boundary data map $\psi$ {\it does not scale} at all.  This leads to
an additional factor $\varepsilon$ in the operator $\Lambda$ determining the
boundary operator $B$ of Eq. (\ref{5}).  
Thus, we define the {\it graded scaling} of
the boundary operator by
\be
B^\varepsilon=\left(\matrix{\Pi^\varepsilon&0\cr
\varepsilon\Lambda^\varepsilon& \II-\Pi^\varepsilon\cr}\right),
\label{34}
\ee
which has the following asymptotic expansion in $\varepsilon$:
\be
B^\varepsilon\sim\sum\limits_{n\ge 0}\varepsilon^{n} B_{(n)},
\label{35}
\ee
where $B_{(n)}$ are first-order tangential operators 
with homogeneous symbols. At zeroth order we have
\be
F_0=-\partial_r^2-\hat\partial^2,
\label{36}
\ee
\be
B_{(0)}=\left(\matrix{\Pi_0&0\cr
\Lambda_0& \II-\Pi_0\cr}\right),
\label{37}
\ee
where 
\be
\hat\partial^2=\hat g^{jk}(\hat x')\hat\partial_{j}
\hat\partial_{k},\qquad
\Lambda_0=\Gamma^j(\hat x')\hat \partial_j,\qquad
\Pi_0=\Pi(\hat x').
\label{38}
\ee
Note that all leading-order operators $F_0$, $B_{(0)}$ and $\Lambda_0$ have
{\it constant} coefficients and, therefore, are very easy to handle.  This
procedure is called sometimes ``{\it freezing the coefficients} of the
differential operator''.

The subsequent strategy is rather simple.  Expand the scaled heat kernel in
$\varepsilon$ and substitute into the scaled version of the equation
(\ref{26}) and of the boundary condition (\ref{28}).  Then, by equating
the terms of the same order in $\varepsilon$ one 
gets an infinite set of recursive equations
which determine all $U_{B(n)}$.  The $U_{\infty(n)}$ are obtained simply by
expanding the scaled version of (\ref{22}) in power series in
$\varepsilon$.

\section{Parametrix: Leading Order}

In this section we determine the parametrix of the heat equation to leading
order, i.e.  $U_0=U_{\infty(0)}+U_{B(0)}$.  As we already outlined above, we fix
a point $\hat x'\in \partial M$ on the boundary and the normal coordinates at
this point (with $\hat g_{ik}(\hat x')=\delta_{ik}$), take the tangent space
$T(\partial M)$ and replace the manifold 
$M$ by $M_0\equiv T(\partial M)
\times{\bf R}_{+}$.  By using the explicit form of the zeroth-order operators
$F_0$, $B_0$ and $\Lambda_0$ given by (\ref{36})--(\ref{38}) we obtain
the equation
\be
\left(\partial_t-\partial_{r}^{2}-\hat\partial^2\right)U_0(t|x,y)
=0,
\label{39}
\ee
and the boundary conditions
\be
\Pi_0U_0(t|x,y)\Big|_{r(x)=0}=0,
\label{40}
\ee
\be
(\II-\Pi_0)\left(\partial_r+i\Gamma^j_0\hat\partial_j\right)
U_0(t|x,y)\Big|_{r(x)=0}=0,
\label{41}
\ee
where $\Pi_0=\Pi(\hat x'), \Gamma^j_0=\Gamma^j(\hat x')$.  Hereafter the
differential operators always act on the first argument of a kernel.  Moreover,
for simplicity of notation, we will denote $\Pi_0$ and $\Gamma_0$ just by $\Pi$
and $\Gamma^j$ and omit the dependence of all geometric objects on $\hat x'$.
To leading order this cannot cause any misunderstanding.  
Furthermore, the heat kernel should be bounded,
\be
\lim_{r(x)\to\infty}U_0(t|x,y)
=\lim_{r(y)\to\infty}U_0(t|x,y)=0 ,
\label{42}
\ee
and symmetric,
\be
U_0(t|x,y)={U_0(t|y,x)}.
\label{43}
\ee

To solve the above boundary-value problem we use the Laplace tarnsform in $t$
and, since it has constant coefficients, the Fourier transform in $(\hat x-\hat
y)$.  Therefore, it reduces to an {\it ordinary} differential equation of second
order in $r$ on $\RR_+$, which can be easily solved taking into account the
boundary conditions at $r=0$ and $r \to \infty$. 
Omitting simple but lengthy calculations we obtain
\be
U_0(t|x,y)=\int\limits_{{\bf R}^{m-1}} 
{d\zeta\over (2\pi)^{m-1}}
\int\limits_{w-i\infty}^{w+i\infty} 
{d\lambda\over 2\pi i}\,
e^{-t\lambda +i\zeta\cdot(\hat x-\hat y)}
G(\lambda|\zeta,r(x),r(y)),
\label{44}
\ee
where $w$ is a negative constant and $G$ is the 
leading-order resolvent kernel in momentum representation. It reads
\bea
&&G(\lambda|\zeta,u,v)
={1\over 2\sqrt{|\zeta|^2-\lambda}}\Bigg\{
\exp\left\{-|u-v|\sqrt{|\zeta|^2-\lambda}\right\}
\nonumber\\
&&
+\left[\II -2\Pi+2i\Gamma\cdot\zeta
\left(\II\sqrt{|\zeta|^2-\lambda}-i\Gamma\cdot\zeta\right)^{-1}\right]
\exp\left[-(u+v)\sqrt{|\zeta|^2-\lambda}\right]
\Bigg\},
\nonumber\\
&&
\label{45}
\eea
where ${\rm Re}\,\sqrt{|\zeta|^2-\lambda}>0$.
Now, by scaling the integration variables $\lambda\to \lambda/t$
and $\zeta\to \zeta/\sqrt t$ and shifting the contour of integration
over $\lambda$ ($w\to w/t$, which can be done because the integrand
is analytic in the left half-plane of $\lambda$) and using the homogeneity
property of the resolvent kernel we obtain immediately
\bea
U_0(t|x,y)&=&(4\pi t)^{-m/2}\int\limits_{{\bf R}^{m-1}} 
{d\zeta\over \pi^{(m-1)/2}}\,
\exp\left\{i\zeta\cdot{(\hat x-\hat y)\over \sqrt t}\right\}
\nonumber\\
&&\times
\int\limits_{w-i\infty}^{w+i\infty} 
{d\lambda\over  i\sqrt \pi}\, e^{-\lambda}\,
G\left(\lambda\Big|\zeta,{r(x)\over\sqrt t},
{r(y)\over \sqrt t}\right).
\label{46}
\eea
Next, let us change the variable $\lambda$ according to $\lambda \equiv
|\zeta|^2+\omega^2$.  In the upper 
half-plane, ${\rm Im}\,\omega>0$, this change
of variables is single-valued and well defined.  Under this change the cut in
the complex plane $\lambda$ along the positive real axis from $|\zeta|^2$ to
$\infty$, i.e.  ${\rm Im}\,\lambda=0,\quad|\zeta|^2<{\rm Re}\, \lambda<\infty$,
is mapped onto the whole real axis ${\rm Im}\,\omega=0,\quad -\infty<{\rm
Re}\,\omega<+\infty$.  The interval ${\rm Im}\,\lambda=0,\quad 0<{\rm Re}\,
\lambda<|\zeta|^2$ on the real axis of $\lambda$ is mapped onto an interval
${\rm Re}\,\omega=0,\quad 0<{\rm Im}\,\omega<|\zeta|$, on the {\it positive
imaginary} axis of $\omega$.  As a function of $\omega$ the resolvent 
$G$ is a meromorphic function in the upper half plane, ${\rm Im}\,\omega>0$,
with simple poles on the interval ${\rm Re}\,\omega=0,\quad 0<{\rm
Im}\,\omega<|\zeta|$, on the imaginary axis.  The contour of integration in the
complex plane of $\omega$ is a hyperbola going from $(e^{i3\pi/4})\infty$
through the point $\omega=\sqrt{|\zeta|^2-w}$ to $(e^{i\pi/4})\infty$.  It can
be deformed to a contour $C$ that comes from $-\infty+i\varepsilon$, encircles
the point $\omega=i|\zeta|$ in the clockwise direction and goes to
$+\infty+i\varepsilon$, where $\varepsilon$ is an infinitesimal positive
parameter.  The contour $C$ does {\it not} cross the interval ${\rm
Re}\,\omega=0,\quad 0<{\rm Im}\,\omega<|\zeta|$, on the imaginary axis and is
{\it above} all the singularities of the resolvent $G$.

After such a transformation we obtain
\bea
U_0(t|x,y)&=&(4\pi t)^{-m/2} 
\int\limits_{{\bf R}^{m-1}} 
{d\zeta\over \pi^{(m-1)/2}}
\exp\left\{-|\zeta|^2
+i\zeta\cdot{(\hat x-\hat y)\over\sqrt t}\right\}
\nonumber\\
&\times&
\int\limits_{C} 
{d\omega\over\sqrt\pi}\, e^{-\omega^2}\,2(-i\omega)\,
G\left(|\zeta|^2+\omega^2\Big|\zeta,
{r(x)\over\sqrt t},{r(y)\over \sqrt t}\right).
\label{47}
\eea
Substituting here $G$ given in Eq. (\ref{45})
and computing Gaussian 
integrals over $\omega$ and $\zeta$
we obtain the ``free'' part
\be
U_{\infty(0)}(t|x,y)=(4\pi t)^{-m/2}
\exp\left(-{|x-y|^2\over 4t}\right)\II ,
\label{48}
\ee
and the boundary part
\bea
U_{B(0)}(t|x,y)&=&(4\pi t)^{-m/2}\Bigg\{
\exp\left\{-{|\hat x-\hat y|^2
+[r(x)+r(y)]^2\over 4t}\right\} (\II-2\Pi)
\nonumber\\
&&
+\Omega(t|x,y)\Bigg\},
\label{49}
\eea
where 
\bea
\Omega(t|x,y)&=&
-2\int\limits_{{\bf R}^{m-1}} {d\zeta\,\over \pi^{(m-1)/2}}\,
\exp\left\{-|\zeta|^2+i\zeta\cdot{(\hat x-\hat y)\over\sqrt t}\right\}
\nonumber\\
&&
\times
\int\limits_C{d\omega\over\sqrt\pi} 
\exp\left\{
-\omega^2+i\omega{[r(x)+r(y)]\over\sqrt t}\right\}
\Gamma\cdot\zeta
(\omega\,\II+\Gamma\cdot\zeta)^{-1}.
\label{50}
\eea
Herefrom we obtain easily the diagonal value of the heat kernel: 
\be
U_{(0)}(t|x,x)=(4\pi t)^{-m/2}\left\{\II+
\exp\left(-{r^2(x)\over t}\right) (\II-2\Pi)
+\Phi\left({r(x)\over\sqrt t}\right)\right\},
\label{51}
\ee
where 
\bea
\Phi(z)&=&
-2\int\limits_{{\bf R}^{m-1}} {d\zeta\,\over \pi^{(m-1)/2}}\,
\int\limits_C{d\omega\over\sqrt\pi}\, 
e^{-|\zeta|^2-\omega^2+2i\omega z}\,
\Gamma\cdot\zeta
(\omega\,\II+\Gamma\cdot\zeta)^{-1}.
\label{52}
\eea
This function can be expressed further as\cite{avresp-cmp98}
\be
\Phi(z)=-2e^{-z^2}\II-2{\partial\over\partial z}\Psi(z),
\label{53}
\ee
where
\be
\Psi(z)=\int\limits_{{\bf R}^{m-1}} {d\zeta \,\over \pi^{(m-1)/2}}\,
\int_0^\infty dp\,
\exp\left\{-|\zeta|^2-(p+z)^2+2ip\Gamma\cdot\zeta\right\}.
\label{54}
\ee
It is not difficult to show that, as $z\to \infty$, 
the functions $\Psi$ and $\Phi$ are {\it exponentially
small}:
\be
\Psi(z)\sim {1\over 2z}e^{-z^2} \left[\II
-{1\over 2z^{2}}(\II+\Gamma^{2})+{\rm O}(z^{-4})\right],
\label{55}
\ee
\be
\Phi(z)\sim {1\over z^2}e^{-z^2} \left[-\Gamma^{2}
+{\rm O}(z^{-2})\right],
\label{56}
\ee
where $\Gamma^{2} \equiv g_{ij}\Gamma^{i}\Gamma^{j}$.
For $z=0$, by using the change $\zeta\to -\zeta$, we obtain
\be
\Psi(0)={\sqrt\pi\over 2}\int\limits_{{\bf R}^{m-1}}
{d\zeta \,\over \pi^{(m-1)/2}}\,
\exp\left\{-|\zeta|^2-(\Gamma\cdot\zeta)^2\right\}.
\label{57}
\ee
Note that this integral converges only when the strong ellipticity condition
$|\zeta|^{2}\II+ (\Gamma\cdot\zeta)^{2} >0$ is satisfied.

\section{$A_{1/2}(F,B)$ Coefficient}

Now we take the diagonal
$U_{(0)}(t|x,x)$ given by (\ref{51}) 
and integrate over the manifold $M$. 
Because the boundary part $U_{B(0)}$ is exponentially small as
$r(x)\to \infty$ we can in fact integrate it only over a narrow
strip near the boundary, when $0<r(x)<\delta$. The difference
is asymptotically small as $t\to 0^{+}$. Doing the change of variables
$z=r/\sqrt t$ we reduce the integration to $0<z<\delta/\sqrt t$.
We see that as $t\to 0^{+}$ we can integrate over $z$ from $0$ to $\infty$.
The error is asymptotically small as $t\to 0^{+}$ and does not contribute to
the asymptotic expansion of the trace of the heat kernel.

Thus, we obtain
\bea
\Tr_{L^2} \exp(-tF)&=&\int_M {\rm dvol}(x)\tr_V 
U_{0}(t|x,x) +O(t^{-m/2+1})
\nonumber\\
&=&
(4\pi t)^{-m/2}\left\{A_0+\sqrt t\, A_{1/2}(F,B)
+O(t)\right\},
\label{58}
\eea
where $A_0$ is given by (\ref{18}) and
\be
A_{1/2}(F,B)=\int\limits_{\partial M}
{\rm dvol}(\hat x)\tr_V b_{1/2} ,
\label{59}
\ee
with
\be
b_{1/2}=-{\sqrt\pi\over 2}(\II+2\Pi)
+2\Psi(0)
\label{60}
\ee
Now, using (\ref{53}) and (\ref{57})
and the fact that $\Psi(\infty)=0$ we get easily
\be
b_{1/2}=
-{\sqrt\pi\over 2}(\II+2\Pi)
+\sqrt{\pi}\,\int\limits_{{\bf R}^{m-1}} 
{d\zeta \,\over \pi^{(m-1)/2}}\,
\exp\left\{-|\zeta|^2-(\Gamma\cdot\zeta)^2\right\}.
\label{61}
\ee
Note again that this integral converges {\it only} when the strong
ellipticity condition is satisfied, 
i.e. $|\zeta|^{2}\II +(\Gamma\cdot\zeta)^2>0$.

Further calculations of general nature, without knowing the algebraic
properties of the matrices $\Gamma^j$, seem to be impossible.
One can, however, evaluate the integral in form of an expansion
in the matrices $\Gamma^i$. The integral
over $\zeta$ becomes Gaussian, which enables one to obtain
\be
b_{1/2}={\sqrt\pi\over 2}\Biggl\{\II-2\Pi
+2\sum_{n\ge 1}
{(-1)^n}{(2n)!\over (n!)^22^{2n}}
\hat g_{i_1i_2}\cdots\hat g_{i_{2n-1}i_{2n}}
\Gamma^{(i_1}\cdots \Gamma^{i_{2n})}\Biggr\}.
\label{62}
\ee

Since our main result (\ref{61}) is rather complicated, we now consider
two particular cases of physical relevance.

\begin{itemize}

\item[I.]
The first non-trivial case is when the 
matrices $\Gamma^i$ form an A\-be\-li\-an al\-ge\-bra, i.e.
\be
[\Gamma^i,\Gamma^j]=0.
\label{63}
\ee
One can then easily compute the integral (\ref{61}) explicitly
and obtain
\be
b_{1/2}={\sqrt\pi\over 2} \left\{-\II-2\Pi
+2(\II+\Gamma^2)^{-1/2}\right\}.
\label{64}
\ee
In the case $\Gamma=0$ we recover the familiar 
result for mixed boundary conditions.\cite{gilkey95,branson90}
In the case $\Pi=0$, this coincides with the result of Ref. 8,
where the authors considered the particular case of
commuting $\Gamma^{i}$ matrices (without noting this explicitly).

\item[II.]
A very important case is when the operator 
$\Lambda$ is a {\it natural} operator
on the boundary.  Since it is of first order it can be only the generalized
Dirac operator.  In this case the matrices $\Gamma^j$ satisfy a Dirac-type
condition
\be
\Gamma^{i}\Gamma^{j}+\Gamma^{j}\Gamma^{i}
=2\,\hat g^{ij}{1\over (m-1)}\Gamma^{2} ,
\label{65}
\ee
which leads to\cite{avresp-cmp98} 
\be
b_{1/2}={\sqrt\pi\over 2} \left\{-\II-2\Pi
+2\left(\II+{1\over (m-1)}\Gamma^2\right)^{-(m-1)/2}\right\}.
\label{66}
\ee
Note that this {\it differs substantially} from the result of Ref. 8,
and shows again that the result of Ref. 8 
applies actually only to the completely Abelian case, when all matrices
$\Gamma^j$ commute.  Note also that, in the most interesting applications (e.g.
in quantum gravity), the matrices $\Gamma^{i}$ do not commute.\cite{avresp97}
The result (\ref{61}), however, is valid in the most general case.
A particular realization of the above situation is the ``pure'' Dirac case
when $\Gamma^2=-\kappa(m-1)(\II-\Pi)$, where $\kappa$ is a constant.
In this case we have \cite{avresp-cmp98} 
\be
b_{1/2}={\sqrt\pi\over 2} \left\{-\Pi
+(\II-\Pi)\left[2(1-\kappa)^{-(m-1)/2}-1\right]\right\}.
\label{67}
\ee

\end{itemize}
Thus, a singularity is found at $\kappa=1$. 
This happens because, for $\kappa=1$, the strong ellipticity
condition is violated (see also Ref. 9).
Indeed, the strong ellipticity condition (\ref{11}), 
\be
(\Gamma\cdot\zeta)^2+|\zeta|^2\II
=|\zeta|^2[\Pi+(1-\kappa)(\II-\Pi)]>0,
\label{68}
\ee
implies in this case $\kappa<1$ (cf. Ref. 9).
This is a general feature of the Gilkey--Smith boundary-value problem: 
the heat-kernel coefficients have singularities when the strong
ellipticity condition is violated.\cite{avresp-cmp98}

\section{Boundary Singularities in the Non-Elliptic Case}

Let us now consider the special case when $V$ is a spin-tensor bundle
and the boundary operator
$\Lambda$ is a {\it natural} operator, i.e. the matrices $\Gamma^j$ form
a representation of ${\rm Spin(m)}$. In other words, the matrices $\Gamma^j$
can be constructed only from natural objects like the metric, the normal,
the Dirac matrices and the frame. 
Then the eigenvalues of the leading symbol of the operator $\Lambda$, i.e.
the matrix $\Gamma\cdot\zeta$, depend only on $|\zeta|$, and hence
are linear in $|\zeta|$:
\be
{\rm spec}\, (\Gamma\cdot\zeta)=
\left\{0,\dots,0,\pm i\nu_{(k)}|\zeta|; \; \; d_{(k)}\right\}
\label{69}
\ee 
where $\nu_{(k)}$ are some {\it positive} constants  
and $d_{(k)}$ are the corresponding 
{\it multiplicities}. Here the index $(k)$ labels all {\it
non-zero} eigenvalues. It is put in round brackets to avoid
any confusion with the boundary coordinate index $j$.
Then the strong ellipticity condition (\ref{11}) implies
\be
0<\nu_{(k)}<1.
\label{70}
\ee

Let us compute the fibre trace of the heat-kernel diagonal.
Taking the trace of (\ref{51}) we get
\be
\tr_V U_0(t|x,x)=(4\pi t)^{-m/2}\left\{c_0
+c_1\exp\left[-{r(x)^2\over t}\right]
+J\left({r\over \sqrt t}\right)\right\},
\label{71}
\ee
where
\be
c_{0} \equiv \tr_V \II, \qquad
c_{1} \equiv \tr_V(\II-2\Pi),
\label{72}
\ee
\bea
J(z) &\equiv& \tr_V \Phi(z)
\nonumber\\
&=&
-4\sum_{(k)}d_{(k)}
\int\limits_{{\bf R}^{m-1}} {d\zeta\,\over \pi^{(m-1)/2}}\,
\int\limits_C{d\omega\over\sqrt\pi}\, 
e^{-|\zeta|^2-\omega^2+2i\omega z}\,
{\nu_{(k)}^2|\zeta|^2\over \omega^2+\nu_{(k)}^2|\zeta|^2}.
\label{73}
\eea
Remember that the contour $C$ lies in the upper half plane:  it comes from
$-\infty+i\varepsilon$, encircles the point $\omega=i|\zeta|$ in the clockwise
direction and goes to $+\infty+i\varepsilon$.  

We want to compute the asymptotics of the parametrix as $r\to 0$ while
$t$ is being fixed. This corresponds to the limit $z\to 0^{+}$.
The first two terms are well defined. The real problem is the asymptotics
of the function $J$ as $z\to 0^+$.
The integral over $\omega$ is
calculated by using the formula
\be
\int\limits_C d\omega\, f(\omega)=
-2\pi i\, {\rm Res}_{\omega=i\nu_{(k)}|\zeta|}\, f(\omega)
+\int\limits_{-\infty}^{\infty}
d\omega\,f(\omega).
\label{74}
\ee
The integrals over $\zeta$ can be reduced to Gaussian integrals by lifting the
denominator in the exponent or by using spherical coordinates.

Note that in the integral over $\omega$ all poles of the integrand lie on
the imaginary axis.  Now, if the strong ellipticity condition (\ref{70}) is
satisfied, i.e. all $\nu_{(k)}<1$, then they do not reach the point $i|\zeta|$.
This is very important.  This simple fact leads then to convergence of the
integral over $\zeta$ and {\it regularity} 
of the limit $z\to 0^{+}$. Therefore, the heat-kernel
diagonal is integrable near the boundary, leading to the asymptotics
obtained in the previous Section.

Let us instead suppose that the {\it strong ellipticity 
condition is violated} in
that there is an eigenvalue $\nu=1$ with a multiplicity $d$.  
As is shown in Ref. 1,
in this case the same procedure leads to a {\it singularity} of the
function $J$ as $z\to 0^{+}$, i.e.
\be
J(z)\sim 2d(m-1)\Gamma\left({m\over 2}\right) \,{1\over z^{m}},
\label{75}
\ee 
and hence to a singularity of the parametrix near the
boundary when $r\to 0$, $t$ being fixed.  Moreover, this singularity is
{\it not integrable}, which means that the $L^2$ trace of the heat kernel,
$\Tr_{L^2}\exp(-tF)$, does not exist at all!  This is also reflected in the fact
that the heat-kernel coefficients $A_{k/2}$ become singular. In other words,
the standard form (\ref{17}) of the asymptotic expansion of the heat kernel
is no longer valid.

The singularity at the point $z=0$ results exactly from the pole at $\omega=i
|\zeta|$. In the strongly elliptic case all poles lie on the positive imaginary
line with ${\rm Im}\,\omega <i|\zeta|$, so that there is a {\it finite} gap
between the pole located at the point with the largest value of the imaginary
part and the point $i|\zeta|$.

In the heat-kernel diagonal there are 
three types of terms now.  The first class
of terms do not vanish exponentially when $r\to \infty$.  Those are the interior
terms.  They give the familiar interior contribution when integrated over a
compact manifold.  The second class of terms are those which are exponentially
small when $r\to \infty$ and when $t\to 0^{+}$.  
These are the boundary terms.  When
integrated over the manifold they produce the boundary terms in the standard
heat-kernel asymptotics. In fact, these terms behave, as $t \rightarrow 0^{+}$,
as distributions near the boundary, so that they give well defined non-vanishing
contributions (in form of integrals over the boundary) when integrated with a
function. In the non-elliptic case we 
have however obtained also a third term. This term has an
{\it unusual non-integrable singularity} at the boundary as $r \rightarrow 0$
(on fixing $t$)
\be
\tr_VU_{0}(t|x,x)\stackrel{r\to 0}{\sim}(4\pi)^{-m/2}
2d(m-1)\Gamma(m/2) {1\over r^{m}}.
\label{76}
\ee
Such a singularity is
{\it non-standard} in that:  i) it does not depend on $t$ and ii)
it is {\it not integrable} over $r$ near the boundary, as $r \rightarrow 0$.
This is a direct consequence of the violation of strong ellipticity.

One can ask:  what if the strong ellipticity condition (\ref{70}) is violated
``strongly'', i.e.  there are some eigenvalues that are {\it larger} than one,
$\nu>1$?  Well, then it is not difficult to see that the integrals (\ref{73})
defining the function $J$ diverge for {\it any} $z$. Thus, in this case the
parametrix itself, not only its functional trace, does not exist at all!

\section{Ellipticity in Linearized Gauge Theories}

In this section we are going to show how the 
Gilkey--Smith boundary-value problem can be formulated
in a general gauge theory, following Ref. 1. A linearized
gauge theory is defined by two vector bundles, $V$ and $G$, such that $\dim
V>\dim G$.  $V$ is the bundle of gauge fields $\varphi\in C^\infty(V,M)$, and
$G$ (usually a group) is the bundle of parameters of gauge transformations
$\epsilon\in C^\infty(G,M)$.  Both bundles $V$ and $G$ are equipped with some
positive-definite metrics here denoted by $E$ and $\gamma$, respectively,
that are Hermitian: $E^{\dag}=E, \gamma^{\dag}=\gamma$,
and with the corresponding natural $L^2$ scalar products
$(,)_V$ and $(,)_G$.

The gauge transformations are described by a {\it first-order} differential
operator $R:\ C^\infty(G,M)\to C^\infty(V,M)$.  We restrict to the most
important case when the second-order operator $L:\ C^\infty(G,M)\to
C^\infty(G,M)$ defined by $L=\bar R R$, where $\bar R=\gamma^{-1}R^{\dag}E$, is
a {\it Laplace-type operator} with a non-degenerate leading symbol
$\sigma_L(L;\xi)=|\xi|^2\II_G$.  This means that $\rank \sigma_L(R)=\dim G$.

The dynamics of gauge fields $\varphi\in C^\infty(V,M)$ at the linearized level
is described by a gauge-invariant and formally self-adjoint {\it second-order}
differential operator $\Delta:\ C^\infty(V,M)\to C^\infty(V,M)$.  It is
gauge-invariant in the sense that its leading symbol is {\it degenerate}
and satisfies the identities
\be
\sigma_L(\Delta)\sigma_L(R) = \sigma_L(\bar R)\sigma_L(\Delta)=0.
\label{77}
\ee
We also assume that ${\rm Ker\,}\sigma_L(\Delta) =\{\sigma_L(R)\epsilon\ |\
\epsilon\in G\}, $ and hence $\rank\sigma_L(\Delta)=\dim V-\dim G$.

Instead of the operator $\Delta$ we introduce another formally self-adjoint
se\-co\-nd-or\-der o\-pe\-ra\-tor $F:  \ C^\infty(V,M)\to C^\infty(V,M)$ by 
\be
F \equiv\Delta+R\bar R.  
\label{78} 
\ee
Here we again restrict ourselves to the most important case when $F$ is also a
{\it Laplace-type} operator with a non-degenerate leading symbol
$\sigma_L(F;\xi)=|\xi|^2\II_V$.  Both restrictions made so far are satisfied in
many interesting examples, like Yang-Mills and Einstein theories (for more
details, see Ref. 1)

In quantum field theory one is interested in the one-loop effective
action which is expressed in terms of the 
functional determinants of the operators $F$ and $L$ by
\be
\Gamma^{(1)}={1\over 2}\log\,{\rm Det}\, F
-\log\,{\rm Det}\,L .
\label{79}
\ee

On manifolds with boundary one has to impose some boundary conditions 
to make these operators self-adjoint and elliptic. 
In gauge theories one tries to choose the 
boundary conditions in a gauge-invariant way.
Interestingly, this requirement fixes completely the form of the
boundary operators associated to the operators $F$ and $L$, respectively.

Let us define restrictions of the leading symbols 
of the operators $R$ and $\Delta$ to the boundary, i.e.
\be
\Pi\equiv \sigma_L(\Delta;N)\Big|_{\partial M} ,
\qquad
\nu\equiv \sigma_L(R;N)\Big|_{\partial M},
\qquad
\mu\equiv \sigma_L(R;\zeta)\Big|_{\partial M}.
\label{80}
\ee
Since $L=\bar R R$ and $F=\Delta+R\bar R$ 
are Laplace-type operators, it follows
that $\bar \nu\nu=\II_G$ and $\Pi=\II_V -\nu\bar\nu$.  Therefore, $\Pi$ is a
self-adjoint projector orthogonal to $\nu$, $\bar\Pi=\Pi$,
$\Pi\nu=\bar\nu\Pi=0$.

The requirement of gauge invariance of the boundary conditions 
determines in an almost unique way that the boundary conditions for
the operator $L$ should be of Dirichlet type, 
\be
\epsilon\Big|_{\partial M}=0,
\label{81}
\ee
and the boundary conditions for the operator $F$ should read
\be
\Pi\varphi\Big|_{\partial M}=0,
\qquad
\bar R\varphi\Big|_{\partial M}=0.
\label{82}
\ee

Since the operator ${\bar R}$ in the boundary conditions (\ref{82}) is a
first-order operator, the set of boundary conditions (\ref{82}) is equivalent
to the general Gilkey--Smith scheme formulated in Sec. 2. Separating the
normal derivative in the operator ${\bar R}$ and denoting by $W_0$ the
restriction of the vector bundle $V$ to the boundary, we find exactly the
Gilkey--Smith boundary conditions (\ref{3}) with the boundary operator $B$ of
the form (\ref{5}) involving a first-order operator $\Lambda:
C^\infty(W_0,\partial M)\to C^\infty(W_0,\partial M)$, the matrices $\Gamma^j$
being of the form
\be
\Gamma^j=-\nu\bar\nu\mu^j\bar\nu.
\label{83}
\ee
These matrices are anti-self-adjoint, $\bar\Gamma^i=-\Gamma^i$, 
and orthogonal to the projector $\Pi$, i.e. $\Pi\Gamma^i=\Gamma^i\Pi=0$.

The condition of strong ellipticity then means that the matrix
$|\zeta|\II-i\Gamma\cdot\zeta=|\zeta|\II+i\nu\bar\nu\mu\bar\nu$
should be positive-definite.
The sufficient condition (\ref{11}) of ellipticity now reads
\be
|\zeta|^2\II+(\Gamma\cdot\zeta)^2=|\zeta|^2\Pi+(\II-\Pi)[|\zeta|^2\II
-\mu\bar\mu](\II-\Pi)>0.
\label{84}
\ee
Since for non-vanishing $\zeta$ the part proportional 
to $\Pi$ is positive-definite,
the condition of strong ellipticity takes the form
\be
(\II-\Pi)[|\zeta|^2\II-\mu\bar\mu](\II-\Pi)>0.
\label{85}
\ee

Thus, the following theorem is found to hold: \cite{avresp-cmp98}

\begin{theorem}

The boundary-value problem $(F,B)$ with the boundary operator
$B$ determined by the boundary conditions (\ref{82}) is
gauge-invariant provided that the boundary operator 
associated to the operator $L$ takes the
Dirichlet form.  Moreover, it is strongly elliptic 
with respect to the cone ${\bf C}
-{\bf R}_{+}$ if and only if the matrix 
$[|\zeta|\II+i\nu\bar\nu\mu\bar\nu]$ is
positive-definite.  A sufficient condition for that reads
\be
(\II-\Pi)[|\zeta|^2\II-\mu\bar\mu](\II-\Pi)>0.
\label{86}
\ee
\end{theorem}

We study some explicit examples of gauge theories, 
including Yang--Mills model and Einstein 
quantum gravity, in Ref. 1 and in another contribution
to this volume.\cite{avrespknox}

\section*{Acknowledgments} The authors are grateful to the organizers of the
conference ``Trends in Mathematical Physics'', in particular, to Vasilios
Alexiades, for their kind support and hospitality extended to us at the
University of Tennessee in Knoxville.  The work of G.E.  has been partially
supported by PRIN97 ``Sintesi''.

\end{document}